# MEDICAL GRAPHS IN PATIENT INFORMATION SYSTEMS IN PRIMARY CARE

Thea Hvalen Thodesen, University of Oslo

Uy Tran, University of Oslo

Jens Kaasbøll, University of Oslo, jensj@ifi.uio.no

Chipo Kanjo, University of Malawi, chipo.kanjo@gmail.com

Tiwonge Manda, University of Malawi, tiomanda@gmail.com

**Abstract:** Graphs are very effective tools in visualizing information and are used in many fields including the medical field. In most developing countries primary care, graphs are used to monitor child growth. These measures are therefore often displayed using line graphs, basing it on three indicators (stunting, underweight and wasting) based on the WHO 2006 Child Growth Standard. Most literature on information visualization of electronic health record data focuses on aggregate data visualization tools. This research therefore, was set out to provide such an overview of requirements for computerized graphs for individual patient data, implemented in a way that all kinds of medical graphs showing the development of medical measures over time can be displayed. This research was interpretive, using a user-centric approach for data collection where interviews and web search was used to ensure that the graphs developed are fit the user requirements. This followed prototype development using one of the three free, open source software libraries for Android that were evaluated. The prototype was then used to refine the user requirements. The health workers interpreted the graphs developed flawlessly.

**Keywords:** Medical graphs, Visualization, Patient information systems

## 1 INTRODUCTION

Graphs are very effective tools in visualizing information and have been used in many fields including the medical field. Visualization allows data to be presented in a quick way whilst lifting out the most important facts. In the health sector for example, health workers need to view how different medical measurements develop over time. In most developing countries primary care, graphs are used to monitor child growth. These measures are therefore often displayed using line graphs, basing it on three indicators (stunting, underweight and wasting) based on the WHO 2006 Child Growth Standard (WHO, 2010). Child linear growth failure is known as stunting and often indicates malnutrition in children (Rutstein, 2000). Many health workers in primary care plot measures into graphs by hand, a process that takes a lot of time. Further, once the manually plotted graphs are well used, it makes the graph difficult to read and use as a tool.

This research was carried out in order to design graphing functionality in the DHIS2 Android Capture App (DHIS2, 2020) for community health workers or clinical work in primary care. The DHIS2 software is used in more than 100 low- and middle-income countries for health management information, disease surveillance and individual patient tracking. Utilization of DHIS2 software in these countries has enabled bulky patient registers to be replaced by mobile phones and laptops. Prior to this study, The DHIS2 Capture App for Android had no graph-feature.

Most literature on information visualization of electronic health record data focuses on aggregate data visualization tools (West et al., 2015). There is hardly literature that gives an overview of medical graphs for individual patient data for primary care. This research therefore, was set out to provide such an overview of requirements for computerised graphs for individual patient data, implemented in a way that all kinds of medical graphs showing the development of medical measures over time can be displayed.

Free, open source software packages for drawing graphs in Android were used. This research evaluated the software libraries based on the requirements. Further, the suitability for medical graphs was evaluated through implementing and testing digital graphs amongst community health workers to gauge the usability and usefulness of the graphs.

## 2   METHODOLOGY

This research was interpretive, using a user-centric approach for data collection. Interpretive methodology positions the sense-making practices of human actors at the center of scientific explanation as they encompass an experience-near orientation that sees human action as meaningful and historically contingent (Bevir and Kedar, 2008). Having solicited the initial requirements for the graphs a prototype was developed and subsequent data was collected using the prototype for refined requirements. That way, the respondents would suggest ways of improving the graph after testing the prototype.

The data was collected in two stages, first data collection was done during the month of November 2019 for two weeks. The second data collection, based on the prototype was done during the month of March 2020.

The main research sites were two primary health centres in Malawi, where some of the authors had a research project on patient information systems (Cunningham et al., 2018). Reproductive health was the dominant activity in the clinics, and treating malaria, TB, diahorrea, HIV/AIDS and other endemic diseases constituted other larger areas of work. The clinics had no medical doctors, but 1-2 nurses/midwives, some clinical workers with 1-2 years of health education, and the majority of staff were community health workers (CHW) with a few months of education.

### 2.1   Finding the variety of graph types

To ensure that appropriate visuals are used to represent the medical data, internet search and interviews with health personnel were carried out to find appropriate variety of graph types.

The interviews followed the snowball strategy. Initially, two clinic managers at the two health facilities were asked about the uses of graphs; the two managers guided the researchers to other personnel using graphs in their daily patient work. Additionally, guidelines and manuals present in the clinics were collected. Much as the primary source for data collection was Malawi, the main researcher, who resides in Norway collected some data from a medical doctor in Norway with experience as a GP.

Using the internet, the web was searched with the terms "graph" or "chart" and the medical areas that were mentioned in the interviews. General phrases like "medical" and "care" were also used." The WHO web-site was also searched with the terms "graph" or "chart."

The variety of graph properties found was summarized in a requirement specification document.

### 2.2   Evaluation of graphing software libraries

Internet search and visit to known software repositories were carried out to find graphing software libraries. The software library was to be used on Android devices in the free, open source software DHIS2. Software that was to be accessible from the programming language Java was selected. The software was tested and assessed according to the requirements. Also, documentation of the software and recommendations from other software developers were included in the evaluation criteria.

## 2.3 Testing and refining the graphs in the patient information system

The evaluations aimed at finding possible ways of improving the software rather than measuring efficiency or effectiveness. Variation in data was therefore more important than consistency. Growth monitoring graphs were selected as the type to be tested, due to its widespread use by the majority of primary health care workers. In total, 29 health informants in two health centres tested the app.

The tests took place at convenient places in or outside of the clinic. Interruptions took place when patients or colleagues needed contact with the health workers. Attempts at measuring time for completing tasks were abandoned due to the unpredictable test situation.

Previous research in rural African communities has shown the preference amongst informants to work together in small groups (Winschiers-Theophilus & Bidwell, 2013), thus we allowed for 1-5 health workers to join around the person holding the tablet. While this setting reduces experimental control, the conversations taking place between the health workers constitute a rich data source.

The cheapest hardware will often be purchased in governmental institutions. To keep the test close to reality, the cheapest devices we also selected for the test, and Huawei was the cheapest brand at the point of purchase. Two tablets were chosen, one 10" tablet and one 5" P10 Lite smartphone to determine whether screen size mattered. The app was installed in both devices.

In the field, the app was first presented to the health workers and they were shown how to add new children and register the measures of the child from a visit. Then charts containing the measures registered were demonstrated. The devices were then handed over to the health workers for them to try using.

Data was collected in two areas.

- The functionality of the graph, its presentation and user interaction.
- Health workers' understanding of the graphs.

Data was collected through observing how they interacted with the app and how they communicated with each other to navigate in the app and enter data. Lastly, they were interviewed and discussed with them about their opinions and what to change or add.

## 3 RESULTS

### 3.1 Graph types

Graphs for monitoring of child growth, weight gain during pregnancy, weight of TB patients, lymphocytes and viral load for HIV, glucose and insulin levels for diabetes, four different measures during deliveries, were observed in the clinics and found through internet searches. In addition, several graphs displaying number of cases, including immunization coverage, were observed in the health centres. Since these graphs concern aggregate, and not data on an individual over time, these aggregate data graphs are outside the scope of this study.

Three graphs span the range of requirements for graph functionality. Graphs for child growth and deliveries were used in the clinics visited. The HIV graph was of the textbook-type. It is likely that similar graphs are found in some primary care clinics.

#### 3.1.1 Child growth

Child growth graphs all have the general form illustrated in Figure 1 below. The age on the x-axis could be 0-2, 0-5 or 5-19 years. WHO has tables of normal values and +1, +2, +3, -1, -2 and -3 standard deviations. The health personnel is to take action if the graph crosses a line.

The WHO tables have values for each day during the age 1-5 and for each month 5-19 years, hence the time scale has a finer granulation for younger children.

In addition to weight-for age, there are also graphs for height-for-age and weight-for-height and different standard values for girls and boys and for premature babies. To calculate the SD for a child, the software has to store the approximately 20000 values per graph and interpolate between the data points.

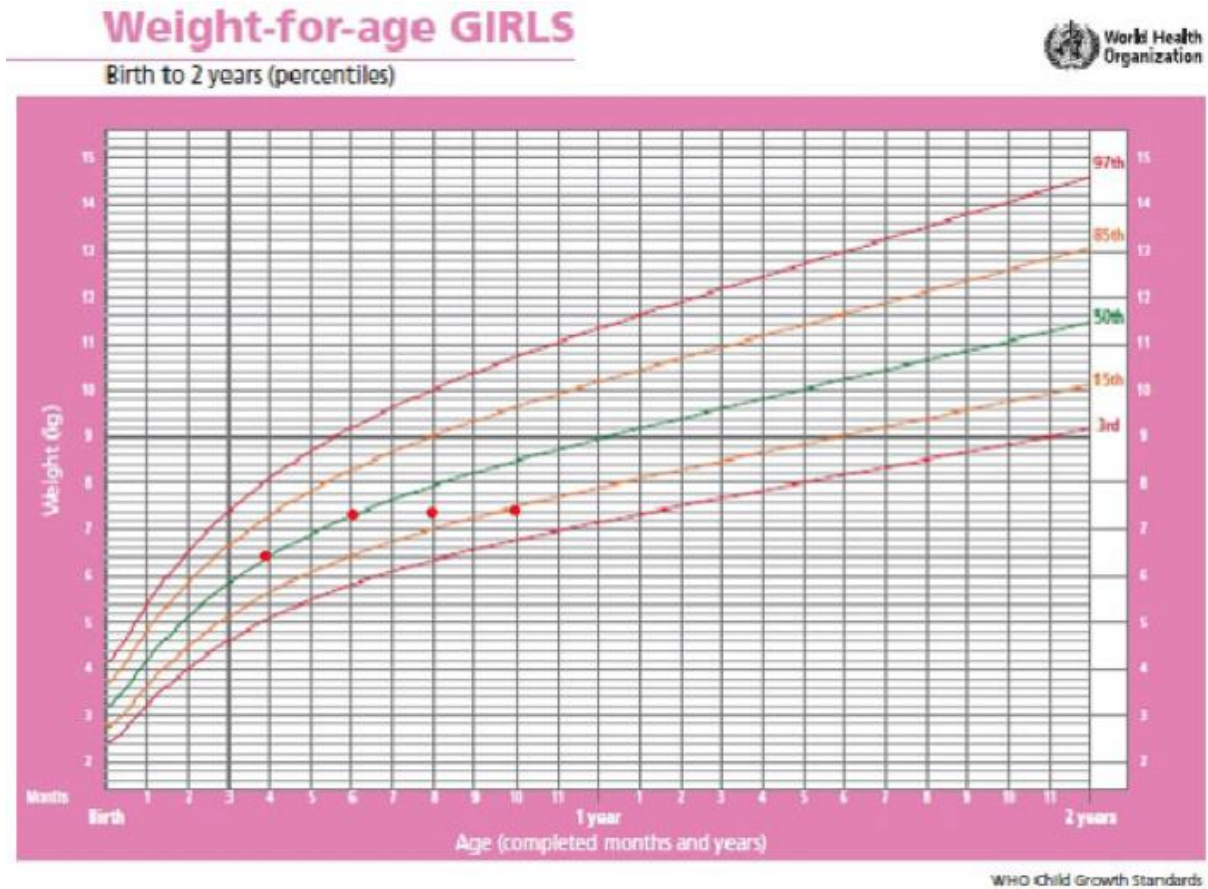

*Figure 1 - Weight for age Girls graph*

Some of the preprinted charts used have colour coding between the SD-lines, eg, red outside of SD 2, yellow between SD1 and SD 2, and green in the middle.

Weight gain during pregnancy had two alert lines above and below the predicted gain (National Research Council, 2010, p. 273). This corresponds to the child growth graph with only the ± 1 SD lines. Heart rate during pregnancy was also recorded, but hot in graphs. However, it could have been displayed in the same type of graph as weight gain.

The graph for monitoring the weight of TB patients only concerns weight gain or loss. Thus it is a much simpler version of the child growth graph.

### 3.1.2 HIV graph

The HIV graph has two series as shown in Figure 2, each presented in separate Y-axes. The lymphocyte Y-axis is linear, while the viral load axis is logarithmic. Its x-axis is more fine-grained to the left than to the right. This shift from high to low granulation along the x-axis is similar to the standard values for growth graphs.

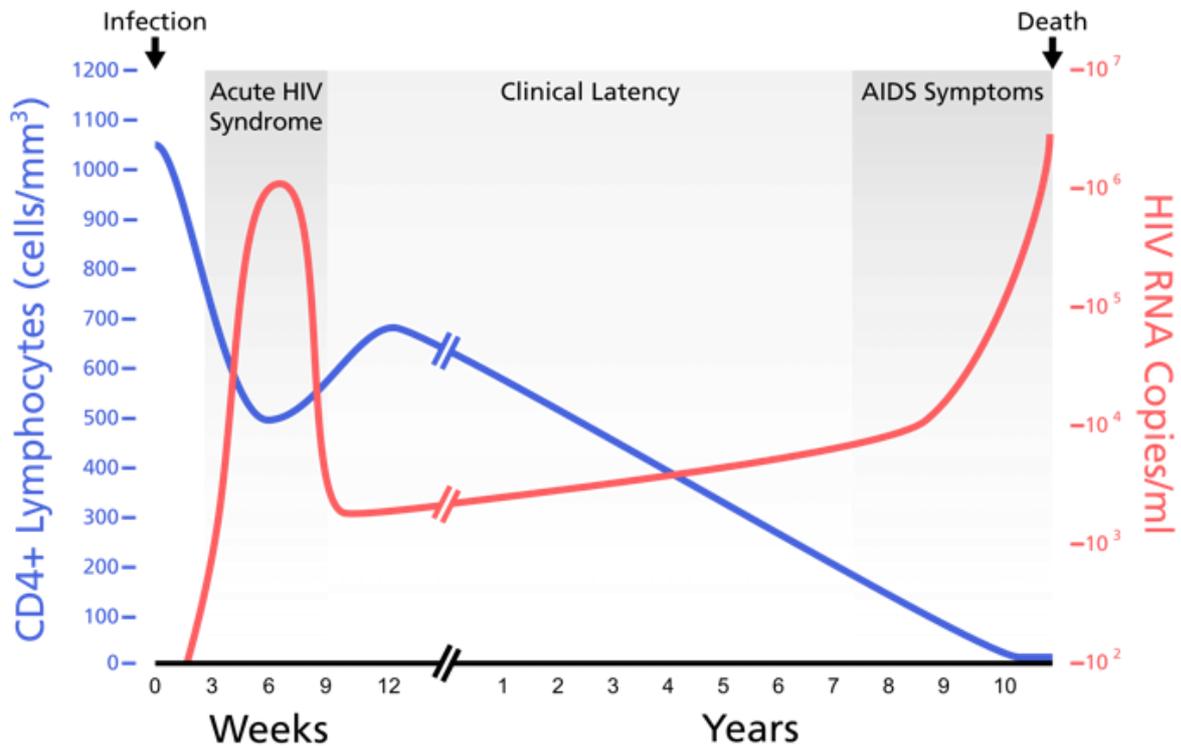

*Figure 2 - HIV Graph*

The diabetes graph also had two lines, each with its own y-axis scale. Since these scales both were linear, the HIV graph requires more functionality in the software.

### 3.1.3 Delivery graph

The delivery graph observed in the clinics had four values per time point at the x-axis; foetal heart rate, cervix opening, descent, and contractions. These values were marked in four different graphs above each other without repairing the x-axis, as shown in Figure 3 below.

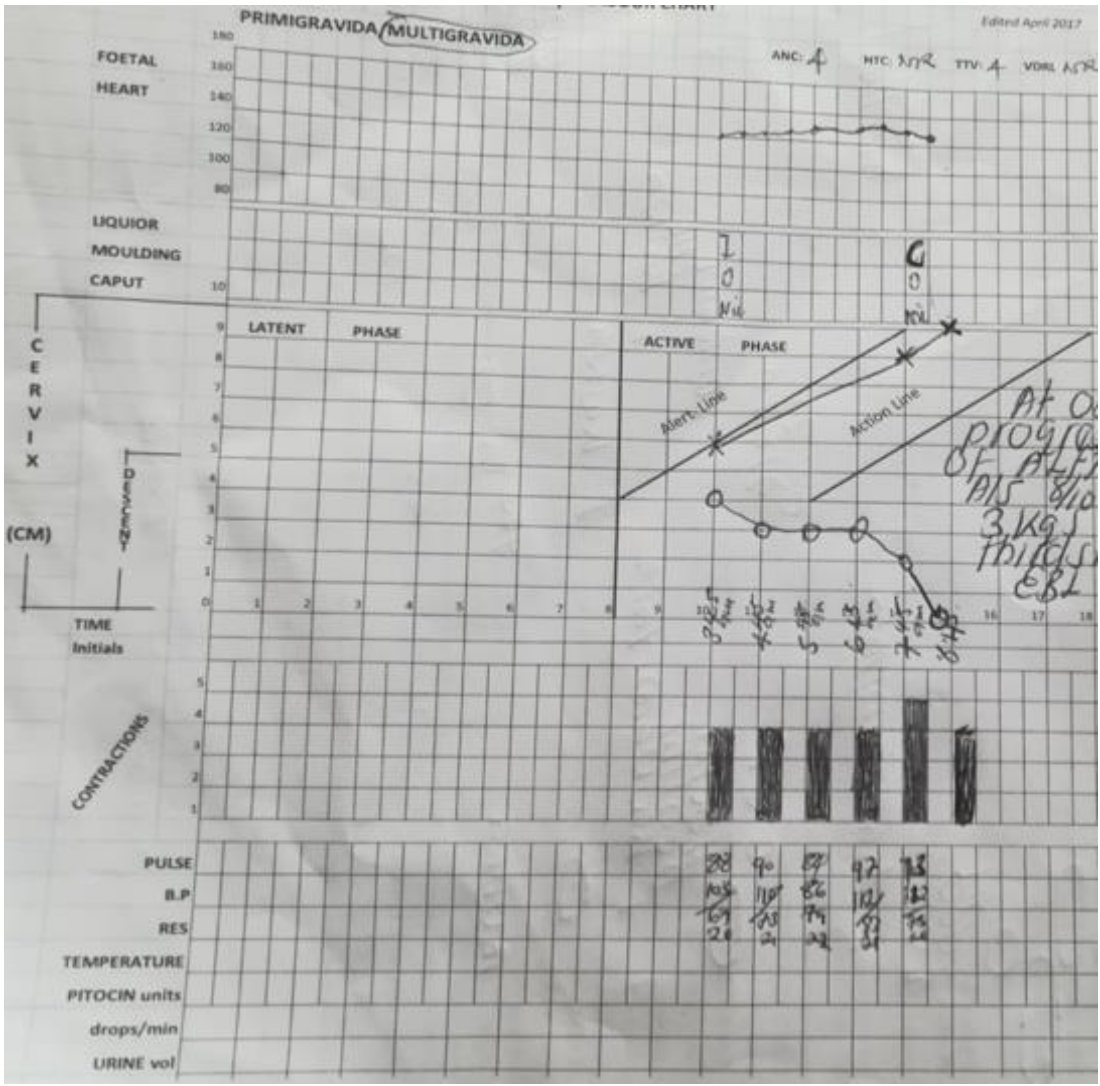

*Figure 3 - Delivery Graph*

The cervix graph has two straight alert and action lines to notify the midwife if the graph crosses the lines.

### 3.1.4 Required functionality

In summary, the requirements for graph functionality were:

1. Any number of series.
2. Linear and logarithmic scale.
3. Double y-axis in one plot.
4. Several charts above each other without repeating the x-axis.
5. Flexible resolution on the x-axis.
6. Standard values, thresholds.
7. Colour between lines.

## 3.2 Code libraries for graphs

We found three open-source, Java libraries for drawing graphs, all licensed under the Apache License 2.0 (Apache Software Foundation, 2004).

**MPAndroidChart** supports several different types of charts like line, bar, pie, bubble, radar, etc. (Jahoda, 2019)**.** It is one of the most discussed online library and is, as far as we could see, the one most people recommend in the different developer forums. It was also recommended by the DHIS2 Android Developers in Spain. It is well documented and easy to work with. From the documentation, it seemed to meet requirements 1-3 and 5-6 . The only thing the documentation did not clarify was whether it would support coloring between two lines in the chart.

• **AndroidPlot** supports several types of charts, both static and dynamic (Fellows, 2019). It also seems to support the same functionality as MPAndroidChart, and is easy to customize. However it is less documented than MPAndroidChart.

• **AChartEngine.** This library (Dromereschi, 2016) has functionality for many chart types, but it does not seem to have been updated for four years.

None of the libraries met requirement 4. To plot several graphs above each other with only one x-axis would require extending the libraries, something which our project did not have the resources to do.

We decided to use the chart library MPAndroidChart, because this was the library that had the best documentation and was easiest to find information and tips on developer forums like StackOverflow and Quora. Another factor in our choice was the recommendation from the DHIS2 Android developer team. Through experimenting, we also found ways to colour the background between two lines in this library, therefore it could meet requirements 1-3 and 5-7.

### 3.3   Coding DHIS2 Android

DHIS2 Android app handles data on individual patients and is configurable to any information needs in primary care. It is connected to a server where data can be shared and also aggregated to produce periodic reports. The app is optimised for slow and intermittent internet connection. It can be used offline, it synchronises data when online, and the data transfer is minimised.

To keep data traffic at a minimum, the graph extension was designed to download the standard data when connected to wifi and only when changes in the standard data was made.

The extension was coded in Java with the Android Studio using the DHIS2 Android Software Development Kit (*DHIS2 Android SDK*, 2020)

When bringing the graph into the user interface, several design decisions had to be made. The group could be shown besides a data input field. This option would make the chart very small, hence difficult to read on a small screen on a phone or tablet. Also, the graphs in the paper documents used by the health workers had a small page on their own, hence they were accustomed to having the graph on a separate sheet. It was therefore decided to make a graph tab in the app, which could display the graphs of the relevant health programme, see Figure 4 below. This screenshot is from a child Outpatient Therapeutic Program, where the user has entered data for weight and height and thereafter pressed the Charts tab. The Weight for Age comes up, and the user can also choose the two other available graphs for this programme; Height for Age or Weight for Height.

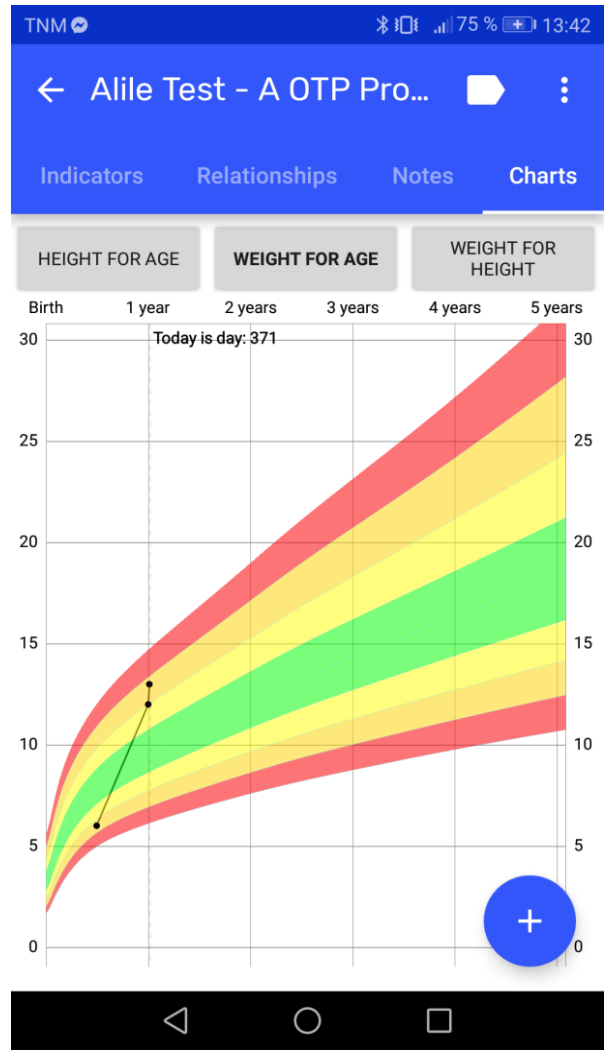

*Figure 4 - OTP Graph*

This project was limited to developing and testing the Android app. To obtain a production system, the configuration functionality of the system has to be extended with the ability to couple specific data fields with the standard value data sets. This extension of the configuration software was outside the scope of the research project, hence the app extension is a prototype.

### 3.4 User experience

We started off by showing the health workers, mainly CHWs, how the app worked:

• How to create a new child

• How to add a new event/visit

• The tab with the different graphs height-for-age, weight-for- age and weight-for-height.

After demonstrating how it worked, we asked the CHWs to create a new child and add a few visits. They found the schemes for Outpatient Therapeutic Program children and started to add a child, some CHWs reading from the schema and others adding data in the app. When the child was added and some visits were registered, they found the tab for graphs.

### 3.4.1 Colour coding

The health workers were given the app with the colours in the figure above. They remarked that the health passports they were used to fill have two colors only green (2 to -2) and yellow (3 to -3), and that the app should be changed accordingly. Since the colouring might differ in other countries, the colours should be configurable in the software.

When they tested the weight for height graph, we encountered a problem. Since the children are visiting every week when enrolled in the Outpatient Therapeutic Program, the plots in the graph are almost on the same spot on the x-axis. It is limited how much weight they gain or lose in 7 days and the height is always the same in this program, as they use the height from admission for all the next visits. They use a table to find the z-score for weight-for-height, instead of a graph. The colors in the table are the same as the colors we used in the weight-for-height, therefore they did not want us to change the colors in this graph.

The health workers were most impressed by the ability to see the z-score relative to the standard deviation ranges, saying

"With this we do not have to use the tables to calculate the z-score"

### 3.4.2 Zooming

They used their fingers trying to zoom in on the graph on the 5" phone screen. "Are you trying to zoom?", we asked. "Yes, is it possible?". We told him it was disabled, because the zooming function collided with the function for swiping between the tabs. It was also hard to move around when the graph was zoomed in. Their comment was that "If zooming was possible, that would be good", this would also result in more space between the plots in the graphs where this was a problem.

The need for zooming was less urgent on the tablet due to the larger screen size.

### 3.4.3 Data storage

The health workers reported that on many occasions mothers lose their child's health passport; where all the data for a child was registered. Some were aware of the server capability offered through a platform like DHIS2, from where the data can be retrieved even when the data entry device is lost:

"Phone might get lost, but the data is not lost, compared to health passport."

### 3.4.4 Calculating therapeutic doses

The health workers wanted to see other data values in addition to the growth graph, and the MUAC measure was central. We therefore included the option for the app to display four data fields above the chart, see Figure 5below. The values in the four data fields change according to the data point pressed in the chart.

The number of rations of Ready-to-Use-Therapeutic Food is determined by the weight of a child, and the health workers would need to look up in a table in order to get that number. The DHIS2 Android app can be configured without coding to calculate such values; the RUTF is shown in the figure below.

The health workers believed that using the mobile phone would be quicker than having to look up RUTF recommendations in a table.

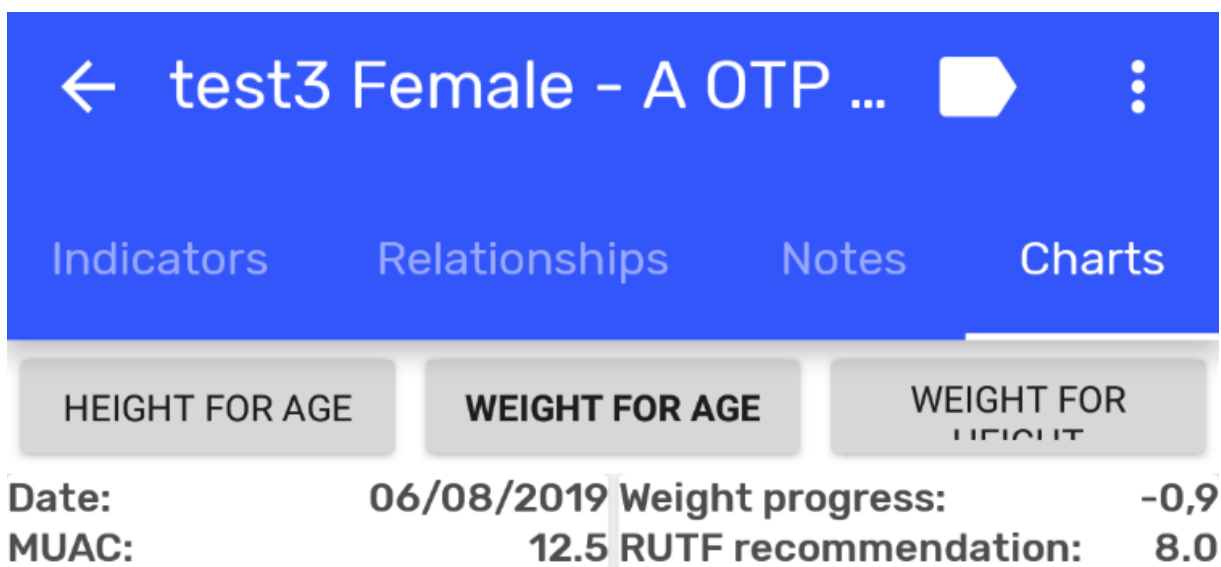

*Figure 5 - Illustration of other data values*

We asked the health worker to plot a child from the Outpatient Therapeutic Program and he could check if the Capture app recommended the same program. The first child he added was admitted to the same nutrition program that the app recommended, the measured MUAC suggested the child should be in the Outpatient Therapeutic Program. "This is very good!". When asked if they thought the messages would be helpful or distracting, they responded that it was "Helpful" and "Not just helpful, they would be very helpful".

### 3.4.5 Satisfaction

When asked how it was to read the charts and how the app was working, he answered "It is easy to use and see what to do. It tells the direct growth of the child." On the question if there is something that should be different, he thinks for a minute, but says it is ok and nothing should be different. We asked if we should remove the weigh-for-height graph, as the plots were difficult to distinguish; "No, keep it. It should be there".

When asked what they think after seeing the app for the first time, they responded that: "It is easy to use" and "Other than fixing the colors, it looks good".

## 3.5 Interpreting graphs and SD-scores

### 3.5.1 Reading growth graphs

Displaying graphs, SD-scores and other calculated values were regarded as improvements by the health workers. Hence, we also wanted to know their ability to relate to the graphs and the numbers in their work.

A simple, first test was showing a health worker four charts of different children and asking him what they would do when this child came to the clinic. We decided to test the weight-for-age graph, since it is always updated in the child's health passport.

The four charts displayed the following:

1. An average child with stable growth. The response was: "This child is normal, we do nothing".
2. Average child with decreasing weight: "This child is normal, but losing weight. We would consult and give guidance to the mother".

3. Child decreasing weight, into the yellow color: "We would admit this child to the Supplementary Feeding Program (SFP) and consult the mother about 6 food groups".
4. Child growth below the yellow field: "This child we would admit to Outpatient Therapeutic Program (OTP) and consult the mother"

These were all adequate interpretations according to their guidelines. After trying out with a few more health workers, the responses were consistently the same, hence the health workers seem to interpret the graph adequately.

### 3.5.2 Looking up in tables

In OTP, the weight-for-length indicator requires the health workers to look up the z-scores in Table 1 and set symbols to denote the inequality. A girl with length 45 cm and weighing 2kg can both be written as < -2 z or > -3 z using the weight-for-length reference table below.

*Table 1*

### Annex 1-9: Weight-for-Length Reference Tables, Birth to 2 Years of Age

| Boys | | | | Length | Girls | | | |
|---|---|---|---|---|---|---|---|---|
| -3 Z | -2 Z | -1 Z | Median | cm | Median | -1 Z | -2 Z | -3 Z |
| 1.9 | 2.0 | 2.2 | 2.4 | 45.0 | 2.5 | 2.3 | 2.1 | 1.9 |
| 1.9 | 2.1 | 2.3 | 2.5 | 45.5 | 2.5 | 2.3 | 2.1 | 2.0 |
| 2.0 | 2.2 | 2.4 | 2.6 | 46.0 | 2.6 | 2.4 | 2.2 | 2.0 |
| 2.1 | 2.3 | 2.5 | 2.7 | 46.5 | 2.7 | 2.5 | 2.3 | 2.1 |
| 2.1 | 2.3 | 2.5 | 2.8 | 47.0 | 2.8 | 2.6 | 2.4 | 2.2 |

According to older and experienced CHWs, most new health workers very often make mistakes when doing this.

When looking at the follow-up schemes for the children admitted to the Outpatient Therapeutic Program, we could see there were some inconsistencies in the calculations of the SD-score. In the figures below, the child (left) has a height of 69 cm, and at 3 visits the weight has been 7.1 kg. In these 3 visits the z-score (WHZ-score in the figure) has been calculated to > -2, < -2, and = -2. Also in another form, for the child at right, the same weight and height has given rise to z-scores both > -2 and < -2.

When questioning them about z-scores, it was clear that some of them found it hard to find the correct z-score, because they changed their opinion and gave different answers.

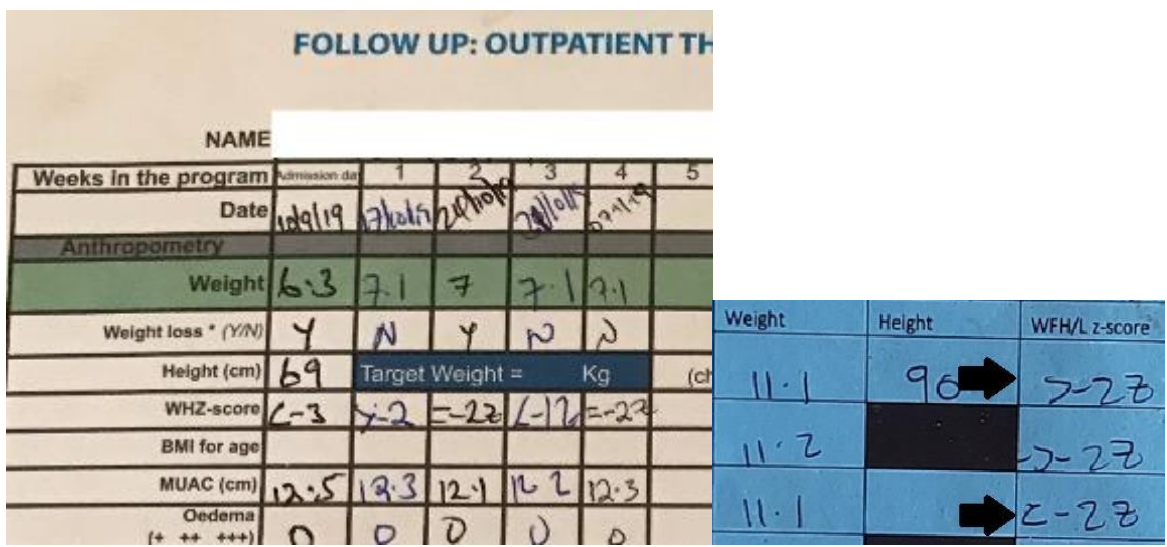

*Figure 6*

Z-scores in the app are viewed when clicking on the plots in the graph. It will show the x-axis and y-axis values of the plot, as well as the z-score. As shown in the form in Figure 6, the health workers are using the greater than and less than symbols when entering the z-score. Our calculation of the z-score displays decimal numbers. This gives a more precise score, but the health workers are not used to z-scores with decimals. When they compared the calculated score from the app and what they found in their reference table, they said "- 1.2 - that is almost correct." Their reference table showed -1. One health worker said, however, "The information is more clear with accurate numbers".

## 4 DISCUSSION

### 4.1 Tables

The health workers interpreted the graphs flawlessly, but when having to look up in Table 1, errors were frequent. In general, the error probability rate for reading a number in a table wrongly is around 1%, and this error rate could increase to 10% when working under moderate stress (Smith, 2017). These general findings apply when a person looks up the figure in a specified row and column. The health workers started with the row which corresponded with the baby's length, thereafter compared the figures in the row with the weight and finally looked up the column heading to find the z-score, hence a more complicated operation than looking up a number in a specific cell. Further, they had to know which column to select when the weight was between two figures in the table. For instance, a 47cm long girl who weighs 2.5kg lies between the figures in the green (2.6) and the yellow (2.4) column. The figures in the column mark the lower bounds of this column, hence the girl is in the yellow zone. Since the table does not indicate whether the figures are lower bounds, middle values or upper bounds of the column, the health workers might fail at making the correct inference for this reason.

The z-scores are marked with < or > symbols in the forms. Amongst middle school students, 30% did not manage to use these symbols correctly for positive numbers (Duru & Koklu, 2011). Also, pre-service teachers have been observed to interpret a comparative symbol (= < >) as a command to perform an arithmetic operation (Ilany & Hassidov, 2018), thus completely misunderstanding the type of symbol. We would therefore expect that people pick the wrong symbol at times.

Ranking negative numbers is another challenge. In s study of children in Sweden learning negative numbers (Kilhamn, 2011, p. 219), one of the high achieving, 12 year old students was given the task to calculate -6-2. His first response was minus 4, with the explanation

" … you have minus 6, and then you take away 2 from that. Then you kind of think that it will be, smaller number, yes?"

He corrected himself to -4, but his initial response assumed -4 < -6. Ordering of negative numbers also came up with large variations in a similar study in Turkey, where the students were given special attention (Altiparmak & Özdoğan, 2010). Hence, a person at any age could make the mistake of mixing up the relationship between the numbers and that of their absolute values.

Health workers filling the forms in figure xxx are thus faced with three challenges; reading the table, using the < and > signs properly, and ordering negative numbers. If there is a 20% chance of failing at each of these three challenges, the result will be random.

The health workers in Malawi had completed high school, and had a few months and up to 2 years of college education. Average percentage of correctly answered items on a statistical numeracy scale for health for people with some college education was 64.5% in the US and 79.2% in Germany. With only high school, the scores were around 10% lower. A study of numeracy amongst patients with university education in the US showed that 40% could not convert a percentage into a proportion of 1000, and 80% failed on the opposite operation (Lipkus et al., 2001).

Based on the three challenges to overcome in the filling of the tables and the frequent math mistakes that educated people do, it is reasonable to assume that many primary care health workers will end up with random use of < and >.

## 4.2   Graphs

In a test of nurses' comprehension of graphs and tables in the US, they performed in the following order; bar graph 88%, table 81%, line graph 77% and spider graph 41% correct responses (Dowding et al., 2018).

In the general population in Germany and the US, the % of correct responses in Table xxx were given to line graph reading tasks (Galesic & Garcia-Retamero, 2011).

| Question no | Task | Overall correct response |
|---|---|---|
| 5 | Reading off a point on a line chart | 83,3% |
| 6 | Comparing slopes of a line at 2 intervals | 71,9% |
| 7 | Projecting future trend from a line chart | 80,5% |
| 12 | Differentiating slope and height of a line | 81,8 |

Hence, around ⅕ of all readings are wrong.

The fluency which the health workers demonstrated with the graphs versus tables contrasts the general findings on graph literacy. There may be several reasons for this.

First, the health worker only needs to see which colour in the area of the data point. Second, no less-than or greater-than comparisons are needed. Third, the health workers do not need to order negative numbers, they just need to see that the data point is below the green level, meaning underweight instead of overweight. Hence, the most difficult operation on paper is to draw the data point and the line. In the app, the drawing is automated, hence this challenge, which could correspond in correctness rate with question 5 in the table is removed.

Another reason for the health workers' ease of dealing with the graphs compared to the general population can be that they are used to one type of line graphs for growth monitoring. Being fluent with this particular graph does not imply a general graph literacy, since specific competencies gained do not easily transfer to the general area.

The graph libraries were used to extend the DHIS2 Android Tracker app, which can store data locally and synchronise with a server when online. To limit data transfer in resource limited places with poor connectivity, standard values were stored on the device and only changed when standard data was changed and when connected to wifi. To make the graph clearly visible, graphs were displayed on a separate tab, where also four essential data fields were displayed.

The 29 health workers appreciated the graph showing the colour codes, such that they did not have to look up in the table. They also wanted to be able to zoom in on the graph, and have the app calculating doses, raising alerts and suggesting treatment programmes.

They interpreted the graphs with ease and chose the appropriate treatment for the children. While people in general make many mistakes when interpreting graphs, the health workers were very familiar with the growth graphs. Also, the colour coding eased the interpretation , such that there was no need for dealing with negative numbers and < and > symbols; the latter would normally cause many mistakes.

While decision support systems have demonstrated mixed results (Bright et al., 2012; Krick et al., 2019), there were cases where the health workers had to consider several conditions simultaneously. The criteria for admission and discharge for Supplementary Feeding Program and Outpatient Therapeutic Program were: a child must have a SD-score between -3 and -2, a MUAC between 11.5 and 12.5 or just been discharged from OTP, while the criteria for Outpatient Therapeutic Program is MUAC less than 11.5, SD-score less than -3 or Oedema +/++ and no other medical complications. The app could be extended without coding to perform such tests and to provide a recommendation. Future research will include such extensions.

# 5    CONCLUSION

We found no prior overview on the types of graphs used for individual patient data in primary care. Through web-search and interviews with health workers, several graphs were found, including graphs for monitoring growth, weight during pregnancy and during TB treatment, HIV, TB, diabetes and deliveries. The following requirements for graphing emerged from these cases:

1. Any number of series.
2. Linear and logarithmic scale.
3. Double y-axis in one plot.
4. Several charts above each other without repeating the x-axis.
5. Flexible resolution on the x-axis.
6. Standard values, thresholds.
7. Colour between lines.

Three free, open source software libraries for Android were evaluated according to their ability to fulfil the seven requirements. None of the libraries fulfilled requirement 4. The MPAndroidChart (Jahoda, 2019) was chosen because it coped with all the other requirements, had the best documentation and was recommended by other developers.